%% file: main.tex
\documentclass[conference,a4paper]{IEEEtran}
\IEEEoverridecommandlockouts
\usepackage{physics}
\usepackage{cite}
\usepackage{amsmath,amssymb,amsfonts}
\usepackage{graphicx}
\usepackage{textcomp}
\usepackage[dvipsnames,table,xcdraw]{xcolor}
\usepackage{cleveref}
\usepackage{multirow}
\usepackage{listings}
\usepackage{float}
\usepackage{fancyhdr}
\usepackage[table]{xcolor}
\usepackage{soul}
\usepackage{xurl} % fix the linebreaks for urls
\usepackage{colortbl}
\usepackage{tikz}

\usepackage{subfig}

\usepackage[ruled]{algorithm2e}

\usepackage{amsthm}

\definecolor{mGreen}{rgb}{0,0.6,0}
\definecolor{mGray}{rgb}{0.5,0.5,0.5}
\definecolor{mPurple}{rgb}{0.58,0,0.82}
\definecolor{mOrange}{rgb}{0.9,0.4,0.1}
\definecolor{backgroundColour}{rgb}{0.95,0.95,0.92}

\lstdefinestyle{CStyle}{
    backgroundcolor=\color{white},   
    commentstyle=\color{mGreen},
    keywordstyle=\color{magenta},
    numberstyle=\tiny\color{mGray},
    stringstyle=\color{mPurple},
    basicstyle=\scriptsize,
    breakatwhitespace=false,         
    breaklines=true,                 
    captionpos=b,                    
    keepspaces=true,                 
    numbers=left,                    
    numbersep=5pt,                  
    showspaces=false,                
    showstringspaces=false,
    showtabs=false,                  
    tabsize=2,
    frame=single,
    rulecolor=\color{black},
    language=C
}

\def\BibTeX{{\rm B\kern-.05em{\sc i\kern-.025em b}\kern-.08em
    T\kern-.1667em\lower.7ex\hbox{E}\kern-.125emX}}

\begin{document}

\title{Envisioning a Safety Island to Enable HPC Devices in Safety-Critical Domains}

\author{Jaume Abella$^1$, Francisco J. Cazorla$^1$, Sergi Alcaide$^1$, Michael Paulitsch$^2$, Yang Peng$^2$, In\^{e}s Pinto Gouveia$^2$
\vspace{0.2cm}\\
$^1$ Barcelona Supercomputing Center, Spain\\	
$^2$ Intel, Germany}

\maketitle

\begin{abstract}
HPC (High Performance Computing) devices increasingly become the only alternative to deliver the performance needed in safety-critical autonomous systems (e.g., autonomous cars, unmanned planes) due to deploying large and powerful multicores along with accelerators such as GPUs. However, the support that those HPC devices offer to realize safety-critical systems on top is heterogeneous.
Safety islands have been devised to be coupled to HPC devices and complement them to meet the safety requirements of an increased set of applications, yet the variety of concepts and realizations is large.

This paper presents our own concept of a safety island with two goals in mind: (1) offering a wide set of features to enable the broadest set of safety applications for each HPC device, and (2) being realized with open source components based on RISC-V ISA to ease its use and adoption. In particular, we present our safety island concept, the key features we foresee it should include, and its potential application beyond safety.
\end{abstract}

\input{1.0.Introduction}

\input{2.0.Background}

\input{3.0.SafetyIsland}

\input{4.0.Components}

\input{5.0.Applications}

\input{6.0.Conclusions}

\section*{Acknowledgements}
{\color{black}
BSC authors contribution is part of the project ({\color{black}ISOLDE}), funded by MCIN/AEI/10.13039/501100011033 and the European Union NextGenerationEU/PRTR, and the European Union's Horizon Europe Programme under project KDT Joint Undertaking (JU) under grant agreement No 101112274. This work has also been partially supported by the Spanish Ministry of Science and Innovation under grant PID2019-107255GB-C21 funded by MCIN/AEI/10.13039/501100011033.
}

\bibliographystyle{IEEEtran}
\bibliography{biblio}

\end{document}

%% file: 1.0.Introduction.tex
\section{Introduction}
\label{sec:intro}

HPC processors can deliver the performance needed for future embedded systems of autonomous cars and aircraft, which will rely heavily on performance-hungry Artificial Intelligence (AI) software, as well as a high number of processes to manage simultaneous events triggered by a plethora of sensors and by timers periodically. However, the support that those devices offer for their use in safety-related applications is different across devices, and so are the guarantees or additional support they need -- if any -- for their use in applications with varying needs in terms of integrity level, fail-safe or fail-operational needs, performance predictability, and time to recover from different types of errors.

In general, HPC processors can be used for fail-safe applications as long as a high-integrity microcontroller unit (MCU) is deployed along with the HPC processor, and the overall system architected so that the MCU can manage all errors affecting both, the HPC processor or the MCU itself, and guide the system to a safe state timely in accordance with application safety requirements. However, if the MCU cannot manage all errors in the HPC processor preserving safety requirements, or if the HPC processor must remain fault-tolerant to meet such safety requirements, then additional support is needed beyond that of the MCU. Such support can be part of the HPC device itself, or be delivered in the form of an enhanced MCU with extended safety capabilities to assist the HPC device.

To overcome the potential limitations that some HPC devices may offer for their use in safety-critical applications, \emph{safety islands} have been proposed recently~\cite{SiemensSI,IntelGo,NVIDIAdrive,NVIDIAdrive2}. While the term \emph{safety island} includes a heterogeneous set of devices, they normally offer two key sets of features: (1) a safe enclave to run safety-related applications, in the same way an MCU does. In fact, an MCU can be regarded as a safety island. However, safety islands offer lower performance than HPC devices, at least for a set of performance-demanding applications, and hence the need for an accompanying HPC device. (2) A safety island also offers specific support to manage an external device that needs to run safety-related functionalities. Such support may include watchdog services, performance monitoring capabilities, ability to initiate test capabilities in the other device or to test it externally, or support to orchestrate diverse redundant execution in the external device to name a few.

When considering the deployment of a safety island along with an HPC device, considerations about integration cannot be neglected. For instance, new cost-efficient chip production and packaging options (multi-die chips -- EMIB\footnote{Embedded Multi-Die Interconnect Bridge}, Foveros~\cite{packagingIntel}) provide a solution to deploy separate dies in the same package, which benefits a number of safety island requirements like fault independence with different dies while preserving high bandwidth communication.

This paper presents our own concept of a safety island, as well as our strategy to realize it based on open source RISC-V~\cite{RISCV} components with the aim of easing its use and adoption. 
In our case, we aim at providing a rich set of safety services to the HPC device to enable the deployment of safety-relevant applications on top with varying safety requirements, despite the potentially limited support that the HPC device may offer natively. Those services foreseen for our safety island include the following:
\begin{itemize}
\item Controllability features to configure the HPC device for fault management and containment, and for providing predictable performance.
\item Observability features for system validation, error diagnosis, and as the basis to build informed safety measures on top.
\item Safety measures to contain and manage errors, and to mitigate abnormal performance conditions.
\end{itemize}

In this paper we provide the following contributions for the design and deployment of our view of a safety island:
\begin{enumerate}
\item We present our concept of a safety island along with examples for its practical realization focusing on open source SoCs based on the RISC-V ISA.
\item We identify existing open source components providing controllability and observability features, and safety measures, appropriate for our Safety Island.
\item We identify further features to be developed in the future to complete and complement our safety island.
\item We devise further applications of the Safety Island beyond safety, such as security, reliability, and power and temperature management. To some extent, the foreseen security extensions are comparable to the ``Security Island'' already offered by some processors (e.g., Arm's TrustZone~\cite{TrustZone}). 
\end{enumerate}

The rest of the paper is organized as follows. Section~\ref{sec:back} provides some background and other solutions comparable or related to the safety island proposed in this paper. Section~\ref{sec:sais} presents our concept of safety island. Section~\ref{sec:comp} describes convenient components for our safety island, whether they already exist or not. Section~\ref{sec:appl} shows how the safety island could be used for other applications. Finally, Section~\ref{sec:concl} summarizes this paper.

%% file: 2.0.Background.tex
\section{Background}
\label{sec:back}

\subsection{Safety-Critical System Development Process}

\begin{figure}[t!]
  \centering
  \includegraphics[width=1.0\columnwidth]{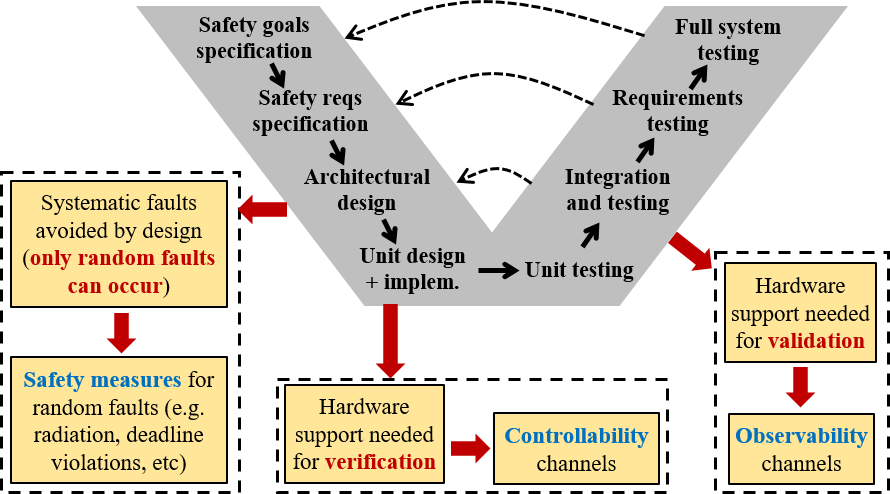}
  \caption{V-model of the development process of a safety-critical system.}
  \label{fig:devproc}
\end{figure}

Domain specific safety standards describe the development process to be followed for safety-critical systems. Examples of such standards include ISO26262~\cite{ISO26262} for the automotive domain, EN5012x~\cite{EN50126,EN50128,EN50129} for the railway domain, and IEC61508~\cite{IEC61508} for industrial systems. Those development processes follow a ``V'' model (see Figure~\ref{fig:devproc}). First, the safety goals of the system need being specified. Safety requirements are obtained out of those safety goals. Then, the architecture of the system is devised ensuring that all safety requirements are mapped to specific items that will have to fulfill them. 

Safety-critical systems are designed to be correct by construction, hence meaning that both, hardware and software, are error free. This is achieved by following specific rules and processes when architecting the system and implementing it. However, random hardware faults due to radiation, sporadic deadline violations, voltage fluctuations, and the like cannot be fully avoided. Hence, safety measures must be included in the system design to manage errors emanating from those faults (e.g., tolerating them, or detecting them and reaching a safe state timely). Once the system has been architected, its different components are implemented adhering to specific constraints dictated by the standards (e.g., avoiding unobvious control flow in the software).

Verification activities for the system architecture and implementation must be included in the process as a way to assess that the design adheres to its requirements and specifications. Controllability means are generally needed along with the design and verification activities to avoid problematic behavior during operation and to enforce specific behavior for verification purposes.

Validation activities (the right part of the ``V'') include testing activities for individual components, as well as for the subsequent integrations with the goal of spotting design errors and, in their absence, gain confidence on the safety of the system. Those testing activities must be as little intrusive as possible to observe the system in operation conditions. Using observability means allow provision of system information without altering the behavior of a system.

Overall, safety-critical systems need to include safety measures, as well as controllability and observability channels to ease their development in accordance with safety standards.

\subsection{Relevant Concepts}

In this paper we build on a number of concepts with broad application across domains, but different names. For the sake of consistency, we resort to automotive naming only (i.e. from ISO26262). 

\textbf{Integrity levels}. There are multiple integrity levels that describe different levels of acceptable risk. They are referred to as Automotive Safety Integrity Levels (ASIL) in the context of automotive, with ASIL-D being the most stringent ASIL, and ASIL-A the least stringent yet with some safety requirements. Finally, an additional level named Quality Managed (QM) is used to refer to items with no safety requirement at all.

\textbf{Dual (DMR) and Triple Modular Redundancy (TMR)}. Systems with the highest integrity levels are often realized building on DMR or TMR for error detection and/or correction. In the particular case of automotive systems, ASIL-D components are generally deployed on Dual-Core Lockstep (DCLS) CPUs. DCLS is an efficient implementation of time and space redundancy for computing cores where the same software runs on two identical cores, but with some staggering (time shift) among them so that a fault affecting both cores simultaneously would cause different errors -- if any -- due to their different state. Hence, errors are detectable by means of simple comparison since redundancy is \emph{diverse}. {\color{black}The potential sources for faults affecting redundant cores, as well as their impact, have been carefully analyzed~\cite{staggeringCauses}, and caveats to deal with the impact of such faults provided~\cite{staggeringCaveats}.}

\textbf{Safe state}. Many systems have a safe state, i.e. a state that, whenever reached, guarantees system safety. For instance, the safe state could consist of a successful transfer of the control to the driver in the case of partially autonomous cars, or stopping the car in a safe location. Reaching such safe state may imply that the corresponding safety system is no longer working. Hence, the safe state may affect the availability of the faulty component or the overall system. {\color{black}Systems lacking a safe state are often referred to as fail-operational systems, whereas those with a safe state are referred to as fail-safe systems.
Note that fault tolerance (for fail-operational systems) and safe states (for fail-safe systems) are intended to be achieved at a given system level, but failures may be allowed at lower levels. For instance, a job of the task controlling the braking system may fail due to a soft error, so that we have a failure at the scope of such microcontroller. However, at a higher level we may have multiple redundant units and a voting system to achieve fault tolerance, or mechanisms to detect the error and potentially enforce a lower driving speed (i.e. a potential safe state) if such fault occurs too often and challenges the timeliness of the braking system.}

\textbf{Fault Tolerant Time Interval (FTTI)}. From the time a fault occurs until it is properly controlled, there is a maximum time affordable in which no hazard can occur. That time interval is referred to as FTTI in automotive, and determines the time needed to recover normal operation, or to reach a safe state where, despite availability may be harmed, safety is preserved.

\subsection{Related Work}

This paper presents a safety island explicitly conceived to provide advanced safety services to an HPC device. However, other works such as that by Siemens~\cite{SiemensSI} already envision their own safety island, and some HPC platforms targeting mainly the automotive domain, such as the Intel Go platform~\cite{IntelGo} and the NVIDIA DRIVE AGX Orin~\cite{NVIDIAdrive,NVIDIAdrive2} already include an automotive ASIL-D compliant microcontroller as a form of ``safety island''. 
In the case of Siemens, their safety island~\cite{SiemensSI} aims at providing a safe enclave for execution and post-mortem test capabilities for the HPC device but, to our knowledge, it lacks advanced observability and controllability features, as well as safety measures, intended to preserve safe operation in the HPC device despite errors, as opposed to our concept.
In the case of the HPC platforms with ASIL-D compliant microcontrollers, such automotive microcontroller is an Infineon AURIX processor in both cases, which is primarily intended to operate as a standalone microcontroller. However, if configured properly, it can provide some of the services provided by the safety island described in this paper, potentially with lower efficiency due to the lack of explicit hardware support for some features, such as multicore interference monitoring and diverse redundancy support for the cores in the HPC device.

For the realm of IoT, the safety island on Intel's Atom\textsuperscript{\textregistered} x6000FE Series~\cite{ElkhartLake} enables functional safety (FuSa) capabilities that detect and attenuate a system fault before it causes or exacerbates further errors (if using the fault $\rightarrow$ error $\rightarrow$ failure sequence terms as described in\cite{avizienis2004basic}). Functionality includes fault monitoring and reporting, on-demand diagnostics measurements, watchdog timers, temperature monitoring, self-diagnostics of the safety island itself and encoding/decoding of communication protocols between the safety island and other external elements.

In the area of security, some Arm processors include the Arm TrustZone~\cite{TrustZone}, which is a form of highly-coupled security island providing a security enclave in Arm processors. Hence, despite with different purposes (security instead of safety), and with a particular degree of coupling (only for a highly-coupled implementation), Arm's TrustZone has some commonalities with the safety island.

%% file: 3.0.SafetyIsland.tex
\section{Architecture of our Safety Island}
\label{sec:sais}

\begin{figure*}[!t]
	\centering
	\includegraphics[width=0.75\textwidth]{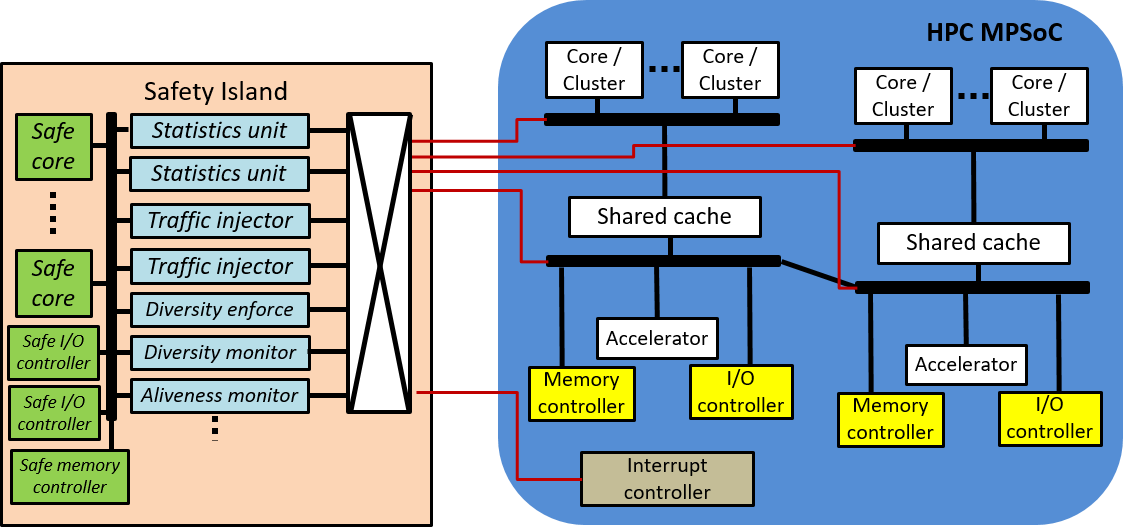}
	\caption{Schematic of our Safety Island.}
	\label{fig:safetyisland}
\end{figure*}

{\color{black}Our goal is presenting a safety island with a number of specific characteristics as follows:
\begin{itemize}
\item Include features for monitoring and controlling the behavior of an HPC device, as well as capabilities to accurately diagnose the cause of any error so that the most effective remedy can be applied.
\item Be suitable for its realization as a chiplet, hence easing integration with COTS HPC devices, and minimizing integration challenges.
\item Be built on open source technologies as much as possible to ease its adoption and extension by the community.
\end{itemize}
}

This section presents our safety island, including key design and integration considerations for its effectiveness.

\subsection{Safety Features of our safety island}

The purpose of our safety island is providing HPC devices with appropriate capabilities to execute performance-demanding applications while preserving their safety requirements. As explained before, safety-related systems require a number of features that can be generally classified into the following categories.

\subsubsection{Controllability Features} 
Controllability features are needed to guarantee that execution conditions for safety-relevant applications are controlled to a sufficient extent (e.g., limiting mixed-criticality interference, guaranteeing some performance levels, etc.). 

\paragraph{Proactive Features}
This type of features includes the ability to restrict the operation of HPC components whenever needed by setting appropriate configurations (e.g., network-on-chip (NoC) policies, cache replacement policies, cache sharing policies, shared resource usage quotas, etc.) so that safety requirements can be implemented through the safety island without introducing disruptive changes in the HPC processor.

\paragraph{Reactive Features}
Alternatively, if some behavior cannot be avoided by means of appropriate configurations, monitoring capabilities are needed to detect misbehavior (e.g., abnormal error detection rates in caches, overutilization of some components, etc.), diagnose the cause, and take actions to mitigate such misbehavior (e.g., stalling the offending process for a while).

\subsubsection{Observability Features}
Observability features are critically important during both system validation and operation. During system validation, they allow collecting detailed information of the behavior of the overall system. The information can be from individual hardware and software components. This information can then be processed offline to detect any misbehavior. Such observability features typically decrease the need for longer test campaigns, since they ease obtaining the evidence; evidence whether some behavior occurs or not. In some circumstances, observability feature may be mandatory to detect certain behavior that is not observable otherwise.
Those features are expected not only to identify specific situations, but to provide enough information to ease diagnosis to mitigate the source of the unexpected situations.

Observability features are fundamental during operation since they can be linked to -- reactive -- controllability features, as indicated before, as well as to safety measures that require detailed diagnostics to guarantee safety without impacting other key metrics such as availability and performance.

The type of information to be observed and how it is collected is highly diverse and must be properly tailored to not incur excessive cost. For instance, one could trace all transactions across two components (e.g., across a shared cache and DRAM memory) or summarize them by counting the amount of data transferred over a period of time. The former may require tracking each individual address accessed along with some information about the transaction (e.g., whether it is a read or store operation, amount of data transferred, etc.), which requires huge storage but provides highly detailed information, whereas the latter provides much less information but only needs few counters to track the amount of data transferred potentially broken down across components originating the transactions (e.g., across cores, accelerators, etc.). 

Overall, it is key to enable the safety island with appropriate support to collect different types of traces, and either output them at high speed, or compress them on-the-fly to generate compressed logs for a ``post-mortem'' analysis. Information to be monitored includes abnormal performance behavior, activity in shared resources (e.g., NoCs, caches and memory controllers), etc. Also, programmable filters to monitor only specific subsets of information are highly convenient.

\subsubsection{Safety Measures}

Random hardware faults and sporadic abnormal performance conditions are generally unavoidable. The safety island must provide capabilities to monitor faults, diagnose the cause, tolerate some of them, and contain the impact of all of them so that software layers can easily and timely manage errors and preserve fault-free operation at all times.

Some safety measures can be implemented by mostly relying on controllability and observability features already described before, but some others may require additional support. For instance, hardware or software monitors may be needed so that, upon the detection of abnormal behavior in the HPC device, specific corrective actions are taken (e.g., resetting specific components, switching to a degraded operation mode, etc.). Since those monitors may have the highest integrity levels, they may require DMR or TMR support in the safety island. Similarly, error detection capabilities for the HPC device may require the safety island to orchestrate some form of diverse and redundant execution on the HPC device regardless of whether the HPC device has specific hardware support for that.

\subsection{Hardware Integration Considerations}

Integration of the safety-relevant features with the HPC device (aka as HPC island) to be mastered is a challenging concern. The higher the coupling between the safety island and the HPC island, the higher the efficiency of the safety island due to having higher controllability and observability, but the lower the modularity since mutual dependencies across the safety and HPC islands would increase. For instance, two extremes of the integration could be as follows:
\begin{itemize}
\item \textbf{Coupled integration}: the islands could be integrated connecting the safety island as a master to the different interconnects in the HPC island (e.g., all AMBA interconnects). This would grant detailed observability and controllability to the safety island, which would have direct information from the different internal interfaces, and could react almost immediately to any predefined event.
On the other hand, such integration would be highly device-specific, hence potentially requiring non-negligible modifications to tailor the safety island to a particular HPC device.
\item \textbf{Loose integration}: the safety island could be encapsulated into a chiplet with standard predefined interfaces agnostic of the particular characteristics of the HPC device to be mastered. This approach would favor portability and modularity, but would be detrimental for observability and controllability purposes since (i) some interfaces may not be directly observable, and (ii) access to the connected interfaces may have higher latency than in the coupled approach, hence increasing reaction time, which ultimately may inhibit the use of some safety measures requiring immediate actions (e.g., to avoid error propagation).
\end{itemize}

When integrating both islands, the safety island and the HPC island, it is critically important to understand what is reachable (and how) through the existing interfaces to program the safety island accordingly (e.g., monitoring modules, traffic injection modules). Note that, even if some parts of the chip are not directly reachable by the safety island, they may still be managed indirectly. For instance, one could inject traffic reaching specific devices attached to not directly reachable interconnects, and observe the latency to obtain a response to guess the amount of load in that interconnect.

{\color{black}
\subsubsection{Chiplets}
While integrating all features in a single chip die generally provides advantages in terms of power and performance, some issues are driving industry towards the use of chiplets instead. In particular, single-chip solutions hinder chip reuse, may lead to lower yield due to the increased number of transistors per chip, and provide diminishing returns in terms of performance as the chip size grows. 

Instead, chiplets provide a number of key advantages, particularly relevant for the safety island: (1) they allow for chip reuse, so the same safety island can be used for different HPC devices. (2) Designs can be specialized for efficiency reasons, hence not needing to build a single chip for all targets. (3) Heterogeneous technologies can be used across different chiplets, hence easing integration. (4) Due to being smaller and simpler devices than monolithic chip solutions, they ease testability. Finally, (5) their smaller size allows increasing yield.

Still, chiplets have to face some challenges that relate to using larger boards to accommodate multiple chiplets, increased latency for chiplet-to-chiplet communication, increased reliability concerns due to the increased number of soldered joints, thermal/mechanical constraints to place multiple chiplets together, and the lack of chiplet-to-chiplet communication standards, although the Universal Chiplet Interconnect Express (UCIe) has recently appeared to mitigate the latter.

\subsubsection{Physical and Logical Integration}
A number of integration aspects related to the physical location of the different chips and the communication interfaces emerge in chiplet-based solutions. For instance, UCIe provides a physical solution to manage the integration of multiple chiplets. However, the particular protocols to be used to communicate chiplets over such communication interface remain to be defined. 

Analogously, whenever multiple chiplets are deployed, the physical location of the different chiplets, including memories, must be carefully laid out to maximize performance, while minimizing power, area and reliability concerns. Also, it must be carefully analyzed which chiplets need to interact. For instance, it is unclear whether all or just a subset of the chiplets need to access main memory, whether it is better deploying memory controllers in all chiplets or concentrating memory access through a single chiplet, whether some specific memory technologies are more convenient than others when using chiplets. Overall, new integration-related challenges emerge when using chiplets. 

Aspects related to the power supply, power domains, and power monitoring must also be taken into account since the safety island is intended to be used for safety critical functionalities. In general, the safety island inherits similar requirements to those of any other safety-relevant microcontroller. Using other chiplets for other devices (e.g., for the HPC device) brings increased costs, as mentioned before, but may ease implementing some safety measures since physical segregation, in general, reduces the number of potential single-point faults.
}

\subsection{System Software Considerations}
A number of observability and controllability features in the safety island may require accessing specific modules in the HPC island such as, memory interfaces to inject traffic, or configuration registers to exercise control on the HPC device. However, existing Memory Management Units (MMUs) or Input/Output MMUs (IOMMUs) may exercise control on the permissions and privileges to access (and potentially modify) different components in the HPC island. Therefore, it becomes critically important that the safety island -- or the system software on its behalf -- is capable of properly configuring permissions and privileges (e.g., at boot time) so that it can perform its work during operation. 
Note that such boot and configuration process is not exempt of security risks, and, hence, is a delicate process that needs to follow appropriate rules for a secure boot and configuration.
Additionally, in order to retain flexibility, namely in regards to fault tolerance and containment, permissions and privileges should as well be configurable at runtime. However, such flexibility can equally incur security risks if the reconfiguration interface represents a single point of failure and can be directly manipulated by exploiting, e.g., a page table vulnerability.

Another software aspect relates to the fact that user level software with safety requirements may need to run on the safety island. For instance, control applications needing native DCLS only available in the safety island may require running on the cores of the safety island. Similarly, some parts of the applications needing to run on the HPC island may also need to run on the safety island. To guarantee a safe environment, virtualization becomes mandatory as well as an appropriate hypervisor or real-time operating system (RTOS) providing partitioning services to those applications, such as fentISS' XtratuM~\cite{xtratum}, SYSGO's PikeOS~\cite{PikeOS}, GMV-Portugal's Air~\cite{GomesOBDP}, Lynx Software Technologies' LynxSecure, Wind River's hypervisor, Green Hill's Integrity, Continental's OSEK VDX, and Erika Enterprise, to name a few. 

\subsection{System Safety Considerations}
The integration of the safety and HPC islands is not exempt of some safety considerations, mostly related to the hardware and software considerations above. A key safety consideration relates to the latency to retrieve information from the HPC island as well as to exercise control. Coupled integrations generally lead to lower latencies, which favor decreased safety risks, as opposed to loose integrations. 

The ability to observe and control the HPC island with low latency is key for a number of safety-related aspects such as fault containment and reaction times at system level. For instance, in the context of automotive safety-relevant systems, an FTTI is defined, as explained before. Exceeding that FTTI implies that hazards can occur potentially violating the safety goals of the system. Therefore, the safety island design and integration must be planned to adhere to safety considerations relevant at hardware, software and system level.

\subsection{Security Considerations}
\label{sec:seccons}

{\color{black}While the goal of the safety island is providing safety capabilities, it must realize some security support to avoid its improper use. Security aspects relevant for the safety island include authentication, permissions and secure boot.

The safety island must realize a secure boot process, in cooperation with software, to guarantee that software executed during booting, as well as drivers loaded, come from legitimate sources. Authentication is also key if any such software requires being updated with updates being initiated by external sources.

During operation, in order to monitor and even control the HPC device, appropriate permissions must be set in the configuration of the HPC device so that safety island actions are allowed. In that context, authentication becomes fundamental since analogous actions triggered by devices other than the safety island must not be allowed. 
Some technologies, such as Intel’s CSME (Converged Security and Management Engine)~\cite{SAI} provide capabilities to authenticate and load firmware into relevant IPs. Such type of technology could be expanded towards use among chiplets, allowing for example the safety island to authenticate itself towards the cores in the HPC device.
}

\subsection{Extendability Considerations}

{\color{black}The use of a safety island, especially if conceived as a separate chiplet, brings opportunities in terms of extensions and updates. For instance, improved safety island designs can be deployed along with a given HPC device without needing to update the latter. Such improvements may come in the form of extensions to enable more efficient means to manage safety aspects (e.g., resorting to hardware support instead of software-only solutions), such as enhanced monitoring units, or more powerful cores in the safety island to perform specific services (e.g., implementing domain-specific ISA extensions in the safety island cores).
Such extensions can be also realized deploying an eFPGA in the safety island so that a firmware update suffices, hence avoiding physical changes.
}

%% file: 4.0.Components.tex
\section{Key Components and Technologies}
\label{sec:comp}

Realizing a safety island requires an SoC capable of executing safety-relevant functionalities, as well as capable of providing safety services to the HPC island. To build a functional and open source safety island, we identify a number of existing and under development components and technologies that need to be consistently integrated to form the safety island. Some of those components and technologies are introduced in this section.

\subsection{Baseline MPSoC}
As part of the H2020 SELENE project, an open source RISC-V based MPSoC suitable for the space, automotive and railway domains has been released~\cite{SELENEpaper}. The SELENE SoC offers a 6-core multicore based on Gaisler's NOEL-V cores~\cite{NOELV} and other GPL IPs~\cite{GRLIB}. Moreover, it includes a wide subset of the IPs described in the remaining of this subsection that make it further appropriate as the starting point to develop a safety island.

{\color{black}
However, there are other alternatives. Unfortunately, high-performance RISC-V cores are mostly proprietary, such as SiFive's P650 and others.
Some open source cores have been recently compared~\cite{dorflinger2021comparative}, including Rocket~\cite{asanovic2016rocket}, BOOM~\cite{celio2019broom}, CVA6~\cite{COREV}, and SHAKTI~\cite{gala2016shakti} C-Class implementations. No core is proven superior to the others in all fronts with varying conclusions for both ASIC and FPGA realizations if we consider performance, power efficiency, area, or maintainability. 
}

\subsection{Multicore Interference Monitors}
A key safety service in multicores relates to monitoring the interference across cores or other type of devices (e.g., accelerators) since such interference may affect real-time guarantees for safety-critical real-time tasks. Recently, the Safe Statistics Unit (SafeSU)~\cite{SafeSU,SafeSU2} has been proposed. It provides capabilities to monitor the traffic in AMBA interfaces such as AHB and AXI4, although its design has been made modular to enable its porting to other interfaces. The SafeSU allows measuring the interference each master device causes on each other device in different interfaces, and has been successfully integrated in the SELENE SoC~\cite{SELENEQoS}.
It remains to be studied how to tailor it to monitor the traffic in remote interconnects rather than those in the safety island itself.

\subsection{Multicore Interference Quotas}
The SafeSU~\cite{SafeSU,SafeSU2} has also been equipped with an interference control mechanism building on its interference monitoring capabilities. In particular, the SafeSU allows programming interference quotas that, upon being exceeded, trigger interrupts that can be immediately captured by the hypervisor or RTOS so that any action needed can be taken, in accordance with system needs (e.g., dropping the offending task, stalling it for a while, increasing QoS guarantees for the offended task). These interrupts have been properly connected to the corresponding interrupt controller at hardware level and successfully captured by the operating system on top, so the integration of the SafeSU with the software layers is simple.

\subsection{Performance Validation}
While monitoring and quota features during operation are key features needed for the design of the system, such system must be thoroughly tested to guarantee that timing overruns will not occur making deadline violation risk residual. Software tests provide limited controllability to exercise all performance corners since multicore interference scenarios can only be induced indirectly and, in some cases, without synchronous control. For instance, some traffic with long bursts can only be produced by devices such as Ethernet ports of the Direct Memory Access (DMA) controller, which are too hard to synchronize with traffic produced by the computing cores. To tackle this issue, the Safe Traffic Injector (SafeTI)~\cite{SafeTI} has been recently proposed. It allows programming arbitrary traffic patterns, including delays between consecutive transactions, fully synchronously, and allowing to generate any type of traffic including read/write, with arbitrary data transfer sizes, with/without burst behavior, etc., including repeated traffic, as well as fixed-size and infinite traffic patterns.
As for the SafeSU, it remains to be studied how to tailor the SafeTI to inject traffic in the HPC island from the safety island.

\subsection{Diverse Redundancy for Cores}
Functionalities with the highest integrity level (e.g., ASIL-D in automotive) require diverse redundancy in several domains, which is efficiently implemented with DCLS. Hence, at least some cores in the safety island need to implement DCLS. The SafeLS realizes DCLS for NOEL-V cores in the SELENE SoC~\cite{SafeLS}. However, as explained before, DCLS is generally expensive if not needed for some tasks since redundant cores are not user visible. Hence, different flavors of diverse redundancy can be deployed providing different tradeoffs, such as allowing cores to be used independently, although failing to provide diverse redundancy for I/O code. This is the case of the Safe Diversity Monitor (SafeDM) module~\cite{SafeDM}, which allows measuring whether diversity exists across two cores. Conversely, the Safe Diversity Enforcement (SafeDE)~\cite{SafeDE} module allows enforcing some time staggering, and hence, diversity across two cores running a task redundantly. The SafeSoftDR software module~\cite{SafeSoftDR} could be used instead since it provides the same functionality as SafeDE in a less efficient manner but without requiring any hardware support. A comparison across the different mechanisms can be found in~\cite{SafeDX}.

Note that, DCLS is intrinsically highly coupled with the redundant cores, and hence, only available for the safety island. Instead, SafeDE, SafeDM and SafeSoftDR can manage diversity for non-DCLS cores. Therefore, they can be tailored to deliver diverse redundancy to cores in the HPC island from the safety island.

\subsection{Diverse Redundancy for Accelerators}
Full redundancy for accelerators such as GPUs is generally not present in HPC devices. Therefore, it is not possible orchestrating diverse redundancy across multiple accelerator instances as done for cores with SafeDE and SafeSoftDR. 

However, accelerators are often highly parallel and offer large internal redundancy this has been leveraged in some works to implement some form of diverse redundancy with appropriate software and hardware support~\cite{divredINTEL,divredNVIDIA}. This type of support can be potentially integrated in the safety island, which can, for instance, offload redundant kernels in a GPU of the HPC island inducing diversity with different means (e.g., intrinsics support, scheduling policy characteristics).

In the context of Deep Neural Networks (DNNs), high -- yet not perfect -- accuracy rates are obtained for processes such as object detection and classification. DNNs often rely on approximation and stochastic behavior, and hence, do not generally require bit-level precision. Instead, high -- yet not full -- precision is wanted at semantic level (e.g., properly detecting and classifying an object) regardless of whether the accuracy is a bit higher or lower. In that context, it is possible deploying lower-cost and approximate (e.g., using lower precision arithmetic) accelerators in the safety island providing diverse redundancy to large and precise accelerators in the HPC island as long as the former are capable of detecting large deviations for the predictions of the latter~\cite{SAURIA}. Such DMR scheme has already been realized in \cite{TRUST}.

\subsection{Watchdogs}
As part of the architectural design of safety-related functionalities, watchdogs are popular since they allow checking the aliveness of specific components. At hardware level, watchdogs are also popular and, in the context of the safety island, they can be deployed to monitor the aliveness of specific components in the safety island as well as in the HPC island (or the complete HPC island). 

Generally, watchdogs are expected to be made sufficiently independent of the item being monitored, e.g., with independent clock and power supply. Hence, this is expected to hold by construction in the case of loose integration of the safety island. However, specific design rules must be followed for both coupled integration of the safety and HPC islands, and watchdogs monitoring components part of the safety island itself.

Watchdogs can monitor clock signals, cycle counters, instruction counters, or time-to-response for some devices. For instance, one could couple a watchdog to the SafeTI so that the latter sends a request requiring response to a specific component in the HPC island, while the watchdog awaits for an answer within a specific time bound. If such response does not arrive timely, the watchdog may raise an interrupt to be captured by the system software in the safety island.

\subsection{Virtualization Extensions}
Hypervisors and RTOSs, often required in safety-critical systems, require appropriate virtualization capabilities to offer partitioning services to the guest operating system. Such virtualization can only be realized if supported by the hardware platform. Hence, virtualization extensions become mandatory for the safety island. For instance, the aforementioned NOEL-V cores in the SELENE SoC implement such extension and have been proven effective to run hypervisors on top, such as fentISS' XtratuM~\cite{SCC,xtratum}.

\subsection{Logging Support}
Most HPC devices include some form of tracing support. However, information traced can be abundant and produced continuously, which, in general, requires a host computer to process it. Such information can be of much use to diagnose the source of some errors, or, at least, to enable reproducibility for diagnostics purposes. 
The safety island can act as such host computer. However, storage capabilities are limited for the safety island (e.g., typically KBs for on-chip storage and MBs for off-chip storage), and hence, either information is dropped by, for instance, retaining the most recent traced information only, or summarized in the form of logs. Retaining recent information can be easily done with trace buffers where information is stored using a FIFO policy. Logging requires, instead, a tradeoff between the details retained and the hardware cost to retain them. The higher the degree of information loss, the lower the cost to store remaining information. For instance, one can track timestamps for specific events, which would require large storage capabilities or restricting trace recording to a limited time window. Alternatively, one could use counters to track occurrences of those events -- potentially broken down across multiple categories, with much lower storage cost, but losing information relative to the timing of those events.
For instance, some authors proposed an error logger for caches tracking error location to diagnose permanent faults~\cite{LogCacheFaults}. 

\subsection{Chiplet Integration Technologies} 
In case of a loose integration with a chiplet-based safety island, Universal Chiplet Interconnect Express (UCIe) comes as the standardizing solution to die-to-die interconnectivity. The layered protocol specifies a die-to-die adapter layer and a protocol layer, the latter supporting PCIe or CXL, with further protocol mappings planned. This requires, however, that communicating chiplets adhere to standards. For instance, UCIe’s specification does not cover packaging/bridging technology used to provide the physical link between chiplets. It is bridge-agnostic, meaning chiplets can be linked via different mechanisms such as fanout bridge, silicon interposers (i.e. 2.5D packaging) or other packaging technologies such as 3D packaging. Nevertheless, standards such as bump pitch, must be taken into account, meaning RISC-V platforms would require dedicated, standardized support for UCIe, which could potentially hinder observability.

In terms of packaging technology, for instance, Intel’s EMIB (Embedded Multi-Die Interconnect Bridge) is a 2.5D packaging technique used to connect dies on the same substrate. 2.5D refers to the integration of dies/chiplets on a substrate using an interposer. It brings specific advantages such as larger die count and larger package configurations, lower cost than full size silicon interposer and support for high data rate signaling between adjacent die.

3D packaging is an alternative to interposers. 3D packaging refers to the direct high-density interconnection of chips through TSV (through-silicon via).
In 3D packaging, chiplets are placed on top of one another instead of horizontally next to one another, forming a 3D structure with each chiplet occupying a layer. Finally, the interposer connects the 3D assembly of the chiplet with the substrate. For instance, Foveros is a high-performance 3D packaging technology.

%% file: 5.0.Applications.tex
\section{Other Applications}
\label{sec:appl}

The safety island can be used, for obvious reasons, in contexts where system requirements include the combination of high-performance needs and safety requirements. However, most of its features can provide other type of services to HPC devices, such as, security and Reliability, Availability and Serviceability (RAS). Therefore, there is a broad area of application for the safety island beyond safety requirements strictly. For the sake of illustration, we introduce the use of the safety island for RAS and security applications in the remaining of this section.

\subsection{RAS}
HPC devices used for servers and supercomputers, to name some application domains, have strict RAS requirements. This relates to the relatively higher criticality of the applications run in those domains when compared with most desktop computers, as well as the much higher exposure to faults due to having very high occupancy and, in the context of supercomputers, typically thousands of computing nodes operating in parallel cooperatively. For instance, in the case of a supercomputer where parallel applications required on average 1,000 CPUs, acceptable failure rates would decrease at least by a factor of 1,000 with respect to single-CPU computers.

Components such as the SafeSU or watchdogs can provide error detection capabilities able to trigger recovery actions at software level. Analogously, logging features can be used to assist recovery by providing relevant information about the error detected, or can also be used to anticipate unrecoverable errors by monitoring recoverable ones. For instance, permanent faults can be detected by diagnosing error location with specific logging capabilities so that appropriate actions (e.g., CPU replacement in a supercomputer) can be taken before permanent faults lead to any unrecoverable error.

\subsection{Security}
It is well-known that all safety-critical systems are also security-critical. This relates to the fact that unintended failures in safety-critical systems could be produced intendedly by an attacker, hence creating at least similar risks. The opposite, instead, is not true, and systems can be security-critical but not safety-critical (e.g., a system managing personal information). 

A subset of the security concerns have analogous behavior to that of the safety concerns, being the only difference whether their root cause is intended (security concerns) or unintended (safety concerns). For instance, abuse in the access to shared resources may be caused by a faulty application or by a malicious attack. In both cases, the SafeSU could be leveraged to, at least, detect the abuse and take corrective actions. Tracing and logging features could be used to discern intended from unintended attacks based on their history or frequency of occurrence, for instance.

Security concerns may often require countermeasures to stop or fool attacks. Components such as the SafeTI may be leveraged for that purpose generating traffic that degrades the ability of the attacker to deduce information from the victim. For instance, in the case of side-channel attacks learning from memory access patterns of the victim, traffic can be injected making the attacker believe such traffic belongs to the victim so that wrong conclusions are reached (e.g., obtaining a wrong key or failing to narrow down enough the possibilities for the victim's key).

In any case, if the safety island is also used for security purposes, a number of additional security-specific technologies may need to be added to the safety (now safety-and-security) island, such as cryptographic accelerators, {\color{black}and means for detecting and defeating attacks.
Note that those security-related features focus on providing security services to the HPC device, whereas those discussed in Section~\ref{sec:seccons} focus on making safety island operation secure.}

%% file: 6.0.Conclusions.tex
\section{Conclusions}
\label{sec:concl}

There is an increasing need for the use of HPC devices in safety-critical systems, but those devices lack enough controllability and observability channels, as well as adequate support to realize key safety measures. Hence, solutions are required to enable the safe use of HPC devices.

This paper presents our concept of a safety island and its main constituents to enable the safe use of HPC devices for safety-critical systems. In particular, we analyzed some key tradeoffs related to the degree of coupling of the safety and HPC islands, identified key components needed or highly convenient to have in the safety island to fulfill its due, and assessed some other types of applications of the safety island with overlapping needs, such as applications with RAS and/or security requirements.